\documentclass[journal]{IEEEtran}
\usepackage{cite}
\usepackage[cmex10]{amsmath}
\usepackage{amssymb}
\usepackage{graphicx}
\usepackage[caption=false,font=footnotesize]{subfig}
\usepackage{color,soul}
\usepackage{setspace}
\usepackage{hyphenat}
\usepackage{subfig}
\usepackage{steinmetz}
\usepackage{bm}
\usepackage{enumitem}
\usepackage{mathrsfs}  
\usepackage{siunitx}
\title{Grid-Forming Loads: Can the loads be in charge of forming the grid in modern power systems?}
\author{Oriol Gomis-Bellmunt,~\IEEEmembership{Fellow,~IEEE,} Saman Dadjo Tavakoli, Vin\'icius A. Lacerda, Eduardo Prieto-Araujo,~\IEEEmembership{Senior Member,~IEEE}}

\begin{document}
\maketitle

\begin{abstract}
Modern power systems are facing the tremendous challenge of integrating vast amounts of variable (non-dispatchable) renewable generation capacity, such as solar photovoltaic or wind power. In this context, the required power system flexibility needs to be allocated in other units that can include energy storage, demand management or providing reserve from renewables  curtailing the output power. The present paper proposes the new concept of grid-forming load, which can be considered a totally flexible concept of demand. The concept is not only ensuring the load is supporting the grid stability by adapting the load to the overall system balancing, but also ensures that the load is actually contributing to form the grid and to provide synchronization power to the overall system. In this sense, the new concept allows running a system powered only by renewables operating at maximum power (or operator defined set-point) in grid-following mode, while the overall system control is ubicated in the demand side. This is an important change of paradigm as it considers that all the flexibility, synchronism and stability provision is on the demand side. This concept can applied either to isolated systems or also to future power systems, where millions of loads steer the power system while the renewables are operating at full power. The paper proposes the concept, suggests possible control implementations of the grid-forming load and analyses the concept in four simulation case studies.
\end{abstract}


\vspace{-3mm}
\section{Introduction} 
Modern power systems are experiencing deep transformations motivated by the massive deployment of renewable energy generation, the irruption of electrical mobility and the digitalization of the energy systems. In traditional power systems, power balancing is provided by conventional power plants (thermal and hydro) that are capable of adjusting the power injection due to the availability of a primary resource, which is stored (in fuel or gas tanks or water reservoirs). Such power plants are interfaced to the power system by means of synchronous generators which provide voltage and frequency regulation, voltage source behavior and inertia to the system \cite{NREL_spin2020}. The whole concept has proven to work very well for a long time due to the flexibility of the generation and has allowed the development of power systems worldwide. 

However, the recent massive deployment of variable (non-dispatchable) power-electronics interfaced renewables, mainly solar photovoltaics (PV) and wind power, has brought a very important challenge and demands to rethink how the whole power system needs to be designed, operated and protected \cite{PES_2018protection,Hatziargyriou_2021}. Solar PV and wind have limited flexibility compared to other technologies which store the prime resource. The flexibility can be provided by operating these non-controllable renewables below the point of maximum power generation in order to have some margin to provide grid support. This allows to involve them in different regulation schemes needed in the power system at different time scales, including slow secondary frequency control \cite{rebel2019}, fast primary frequency support \cite{vaha2015,roge2007,Margaris2012} or grid-forming provision \cite{NREL_GFOR_2020}. These services are considered fundamental for modern power system with massive penetration of renewables, as these renewables are meant to replace conventional power plants \cite{EC_2009,EC_2020}. The grid-forming provision and voltage source behaviour are especially important as the penetration of synchronous machines is being reduced and there is a need for units to form the grid and ensure the stability of the overall system \cite{Matevosyan2019,ENTSOE_2020}.

Regarding the need for flexibility, alternative technologies are emerging as possible providers. The main ones are energy storage systems and controllable demand. Energy storage systems are proposed to be used in multiple grid services at different time scales \cite{Masuta2012} and they are also considered suitable for being operated in grid-forming mode \cite{OSMOSE_2022}. 

As far as controllable demand is concerned, it has the main advantage (compared to energy storage) that it does not need additional investments as the loads need to be supplied anyway. Controllable demand (demand-side flexibility) is suitable in loads which can be deferred or adjusted depending on the system needs. Demand-side flexibility is defined in all time frames: from very fast loads that can respond to the disturbances in sub-second, to very slow loads which change their demands based on a day-ahead forecast for the purpose of energy balancing~\cite{iren2019,anwa2019,sode2018}. Loads with fast demand response have been utilized for various grid services. In~\cite{Zhao2018,zhao2011,doug2013}, control systems of thermostatically controlled loads, such as space heaters and electric water
heaters, are modified to provide frequency support in response to frequency events in the grid. Such frequency support is due to the ability of thermostatically controlled loads to quickly adjust their power consumption based on the frequency variations in the grid. Electric vehicles have been also identified as potential candidates for grid's frequency regulation~\cite{Kaur2018,kemp2008,Han2010}. The demand flexibility of electric vehicles is originated from their large charging and discharging capacities\cite{Kaur2018}, which is exploited by introducing proper control schemes. In case the grid is dominated by industrial drives (such as the case in some medium-voltage microgrids), those drives can provide flexibility to the grid if the electric motors are allowed to operate in different load factors~\cite{chak2017,ryan2021}. More recently, hydrogen electrolyzers gained  attention to be used as flexible loads. An electrolyzer can change its power consumption in fraction of a second to adjust its hydrogen production and store it in a large tank. This demand flexibility is explored in~\cite{dozi2021} to provide fast frequency response to power system. In addition to those examples, there are many other domestic and industrial loads \cite{Babahajiani2016,Guo2020}, whose flexible demands are used for several grid services \cite{Carne2018,Abraham2018}. 

Many of modern AC and DC loads, such as those mentioned above, are interconnected to AC grid via power electronic converters (diode rectifiers or Voltage Source Converters (VSC)) . When VSCs are used, they operate in grid-following mode. A power set-point, which is the power consumption of the load, is tracked by the control loops of VSC. To take advantage of demand flexibility, the power set-point is made a function of the grid's frequency (commonly via a frequency-power droop), so that during under/over frequency events, the power consumption is automatically decreased/increased to minimize frequency deviation in the grid. It should be noted that a VSC operates as a flexible current source in grid-following mode~\cite{yito2022}. The magnitude and angle of such current source are flexibly varied via its control loops to give support (i.e., minimize deviations) to grid's frequency and voltage during transients. Even without an interfacing VSC, an extra power electronic device called as electric spring can be added in series with non-responsive load to make them flexible~\cite{hui2012}. However, to the best of our knowledge, grid-forming capability has not been studied for demand flexibility provision. A major advantage of a load operating in grid-forming mode is that the load provides flexible voltage source behaviour to the grid, as opposed to the current source behaviour in grid-following mode, and actively forms (i.e., directly controls) the voltage and frequency rather than only supporting them. 



This paper proposes a new concept of grid-forming load, which can contribute to the deployment of a future power system fully dominated by renewables. The concept starts on existing proposals of demand side flexibility provision and suggests to bring the flexibility to the extreme, providing grid-forming capability from the load side. This has the main implication of freeing (totally or partially) the renewable power plants to provide this service, thus allowing them to operate all the time at full power following the point of maximum power (no need to keep any reserve margin) or alternatively, if the system requires it, at operator defined set-point. The extreme development of the proposed concept would make possible to have millions of units forming the grid in the future power system with 100\% renewables, where the renewable power plants are operated at full power (or desired set-point). Naturally, this would depend on the limits of flexibility and also on the instantaneous generation available and it would be logical to implement some limitations in the renewable power plant. 

The remainder of the paper is organized as follows. Section~\ref{sec_concept} introduces the grid-forming load concept. Section~\ref{secener} analyzes the energy balancing aspects. Section~\ref{sec_controlmodes} further details the operation and controls models of the grid-forming load. Section~\ref{sec_casestudies} demonstrates the feasibility of the proposed concept by conducting four different case studies in Matlab Simulink. Finally, the conclusions are presented in Section~\ref{sec_conclusions}.

\vspace{-3mm}
\section{Grid-Forming Load (GFM-L)}\label{sec_concept}

Multiple different loads can be subjected to implement the grid-forming (GFM) load concept. The concept is appropriated for loads which can be flexibility operated around an operation point which can be scheduled (by the user or the grid operator). The loads will require to have a VSC interfacing them to the grid. It is important to remark that this is not the case for many loads as typical drives equipped with diode rectifiers, so the proposed concept would bring the need to redesign some loads. It is also important to remark that GFM-L will always require generators in the power system providing sufficient generation. These generators can be installed in  variable renewable-based power plants (solar PV or wind) or dispatchable power plants (like the synchronous generator based conventional hydro or thermal power plants).

\vspace{-3mm}
\subsection{Grid-forming Load (GFM-L) Arrangements and Examples}

\begin{figure}[htbp]
\centering
\includegraphics[width=2.8in]{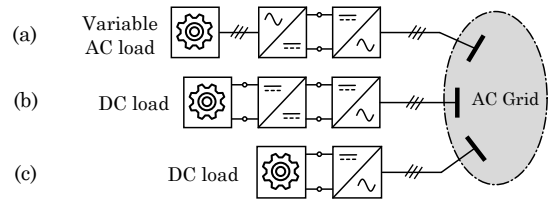}
\caption{Possible topologies of GFM-L.}
\label{GFM_Loads_types}
\end{figure}

\figurename{~\ref{GFM_Loads_types}} shows examples of possible arrangements of the grid forming loads. \figurename{~\ref{GFM_Loads_types}(a)} and \figurename{~\ref{GFM_Loads_types}(b)} are examples of AC loads with a double-stage conversion with an intermediate DC bus, whose voltage is regulated by the load-side converter (DC-AC converter in \figurename{~\ref{GFM_Loads_types}(a)} and DC-DC converter in \figurename{~\ref{GFM_Loads_types}(b)}). In both cases, the grid-side converter provides the GFM capability to the AC grid, while regulating active and reactive power around the set-points, $P^*$ and $Q^*$, respectively. \figurename{~\ref{GFM_Loads_types}(c)} shows the example of a single-stage converter. In this case, the VSC  provides the GFM capability to the AC grid,  regulating active and reactive power around the set-points, and the DC bus voltage (DC load voltage) depends on the equilibrium point. In this last case, it is important to set operation limits carefully to make sure that the DC-bus voltage is within the appropriate boundaries to modulate the voltage in the AC side and operate the converter correctly. This implies a serious limitation and can make this option not suitable.

The concept of GFM-L is suitable for loads which are flexible in nature and can adapt the operating point depending on the systems needs. Some examples of potential GFM-L are listed below:
\begin{itemize}[leftmargin=*]
\item	Resistive loads in electric heaters (or other resistive loads). The resistance can be connected in DC adopting the configuration presented in \figurename{~\ref{GFM_Loads_types}(b)} or \figurename{~\ref{GFM_Loads_types}(c)}, but the option of \figurename{~\ref{GFM_Loads_types}(b)} is preferred to ensure a better operation (regulated DC voltage) of the converter. Heating applications are normally flexible and the load can be managed depending on the system needs. The voltage and current in the resistance will depend on the reference active power, $P^*$, but it will be able to vary around it depending on the GFM control law. This will be possible allowing the load-side converter to regulate the DC bus voltage by changing the voltage (or current) in the resistance. The grid-side converter will implement the GFM control law considering the desired $P^*$. 
\item	Industrial drives. Most current industrial drives have a diode rectifier, a DC bus and an inverter connected to the motor which allows variable frequency  operation and torque (and power) control. To implement a grid-forming load, a back to back scheme will be needed (\figurename{~\ref{GFM_Loads_types}(a)}), including two VSCs. The grid-side converter will be in charge of implementing the grid-forming law while the machine-side converter will control the DC bus voltage. The load power (or speed or torque) will be determined by the grid-side converter setpoint P*. This will be applicable in some loads which can tolerate changes in the power (or torque or speed) but it will not be possible in all loads. 
\item	Electric vehicle (EV) chargers. EV chargers are a flexible load in many circumstances (as users can leave the car parked several hours and they only need to be sure that whenever they need the car, it will be fully charged, regardless when and how the EV will be charged). An exception are fast-chargers which have reduced flexibility if the user needs to charge the vehicle at a specific time. The GFM approach in EV chargers will be very similar to the one described for resistive loads, with a GFM law implemented in the grid-side converter around the charging set-point, $P^*$, and the DC-DC charger is responsible to regulate the DC bus voltage by adjusting the EV charging current. 
\item	Hydrogen (H2) electrolyzers. H2 electrolyzers are emerging as a flexible load which can be successfully combined with renewables to provide flexibility to the network and produce H2 for further applications like hydrogen based vehicles or industry application of hydrogen. 
\end{itemize}

\vspace{-3mm}
\subsection{Advantages of Grid Forming Loads}

\begin{figure*}[htbp]
\centering
\includegraphics[width=0.85\linewidth]{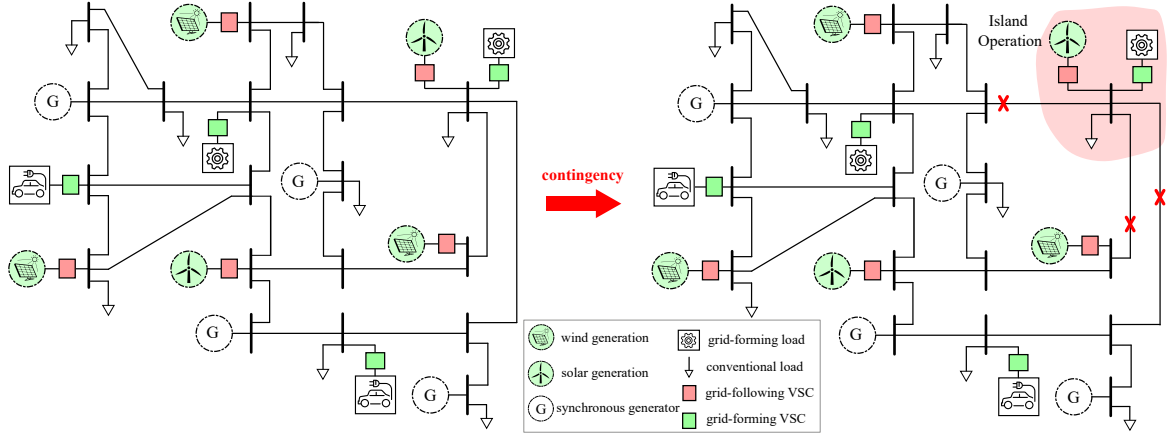}
\caption{Example power system with high penetration of renewables, wherein grid-forming loads support the grid}
\label{sys1}
\end{figure*}

The key advantage of the GFM-L is that it can be defined as a source of extreme flexibility for the power system. The GFM-L concept allows to adopt all the requirements of flexible loads which have been widely discussed, and in addition, it allows to provide all the key advantages of GFM converters:
\begin{itemize}[leftmargin=*]
\item Provide stability to the network,  allowing to regulate and stabilize frequency and voltage. Frequency and voltage regulation can be provided in addition to the grid-forming functionality.
\item Operate as a voltage source to contribute to form the grid.
\item Provide synthetic inertia to the system by setting appropriately the control parameters of the GFM controllers.
\item Reduce the requirements on the generation or storage in the rest of the system. If the flexibility and GFM capability can come from the loads, there is less need to install energy storage and the renewable generation can operate at the point of maximum power (or operator defined set-point) without needed to have a reserve margin for grid support.
\item Increase the system resiliency by allowing islanded operation of a segmented system including some renewable energy sources operating at maximum power (or operator defined set-point). For example, an industry with solar or wind power installed can keep operating when it is isolated from the main grid, operating the renewables at maximum power and steering the grid from the load-side. \figurename{~\ref{sys1}} includes an example of the resiliency. In this example, the system, in normal conditions, combines synchronous generation (providing GFM capability) with grid-following (GFL) VSC in the renewables and GFM-L supporting the grid. A severe contingency can provoke separation of a segment of the network as it is shown on the right of \figurename{~\ref{sys1}}. In this case, the GFM-L keeps forming the grid and adapting the load to the generation available in the wind power plant.
\item  Grid-forming load concept can also be effectively integrated with synchronous generator based power plants.
\end{itemize}

The main limitation of this concept is that a flexible load is required. So, the GFM-L concept can be only implemented for flexible loads. 

\vspace{-3mm}
\subsection{Possible Visions on The Role of GFM-L in Future Power Systems}

The concept can be adopted for future visions on the evolution of large power systems based on variable renewable energy (solar PV and wind). Considering a future power system with very high penetration of variable renewable energy, it is needed to have several units providing GFM capability to the system \cite{Matevosyan2019} which for solar PV or wind implies operating with some margin of reserve \cite{Diaz2014} and/or having some energy-storage units in the GFM mode. This has the implication of wasting a substantial amount of renewable power which could be injected to the grid or require large investments in energy-storage equipment. In this context, the GFM load could be massively adopted and free solar PV and wind plants of providing GFM control, bringing this responsibility to the loads. This is a huge change of paradigm in the power industry. Therefore, this paper intends to highlight that this is a possible development for future power systems. In this new paradigm, we could even have an entire large scale power system running without any single generator forming the grid. All the solar PV and wind power plants would be operating in GFL mode, injecting maximum power (or operator defined set-point) to the grid. The GFM loads would managing the GFM capability and securing the stable operation of the grid. This could be implemented with thousand or millions of distributed loads coordinated by means of a smart grid. Although the communication systems would allow (as it does for the current power system) an optimal operation of the system, the system could still operate safely (with some limitations) when loosing communications or when the grid is segmented during extreme failures.

\begin{figure}[htbp]
\centering
\includegraphics[width=0.85\linewidth]{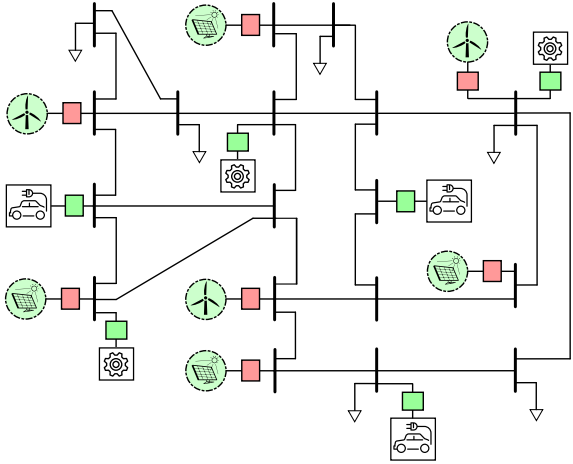}
\caption{Example power system with extremely high penetration of renewables, wherein grid-forming loads form the grid (without synchronous generators)}
\label{sys2}
\end{figure}

\figurename{~\ref{sys2}} illustrates an example system which could be designed to operate solely with GFM loads proving the GFM capability and using several solar PV and wind power plants to inject power in GFL mode. It is important to remark that this is an extreme case, which is relevant to think about the proposed concept, but in practise future power system will often include synchronous generation which could share the GFM role with the proposed GFM-L.  

\vspace{-3mm}
\subsection{Integration of GFM Loads in Future Power Systems}

The previous Subsection was focusing on vision on future power system where the grid-forming loads could be the main or only sources of synchronization and grid forming functionality. This vision  addresses one possible future scenario (which is helpful bringing the concept to the extreme) but there are other which are possible as well. 

Future scenarios can include the integration of multiple GFM units, which can be installed as well in other generation or storage energy sources, while also including power plants based on synchronous generators. The present paper introduces the concept which can be applied in multiple possible schemes. 

In any case, the possibility of using GFM-L in power systems needs to be always considering an intrinsic limitation. The limitation is based on the need to have sufficient variable generation in the overall system in order to make sure that the GFM-L can adjust the load without reducing the load below the minimum threshold. In other words, the GFM-L can form the grid and regulate voltage and frequency if there is sufficient power in the system that can be absorbed by the GFM-L.

\vspace{-3mm}
\section{Energy Balancing Aspects}\label{secener}
\begin{figure*}[htbp]
\centering
\includegraphics[width=0.87\linewidth]{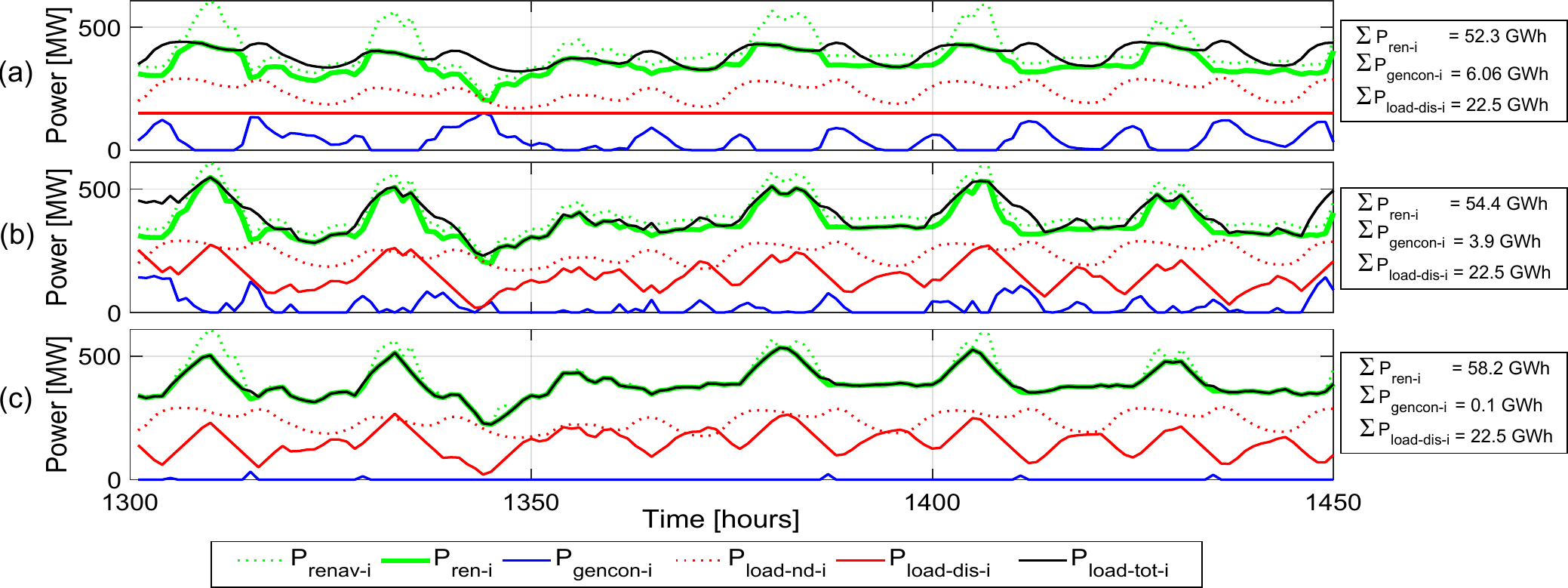}
\caption{Energy balancing analysis of the options A, B and C}
\label{ener1}
\end{figure*}

In the context of modern power systems, where there is a growing penetration of renewables, the responsability of forming the grid need to be ensured without relying only on conventional power plants. Grid-forming capability can be allocated in renewable power plants (with appropiate reserve by curtailment), energy storage units (which have a significant cost) or GFM-L (as suggested in this work). Therefore, the GFM-L concept has important implications in energy balancing of the system. Considering that some units in the system have to provide grid-forming functionality, the GFM-L allows to free other resources from providing it. The GFM functionality in renewables is associated to operate with reserve at the expense of curtailing the resource. A brief example analysis is included in this Section to illustrate this. A system as the one shown in \figurename{~\ref{sys1}} is considered. The system is analyzed for $h$ hours where the available renewable resource is defined as $P_{renav-i}$ for each hour $i \in (1..h)$. The renewable power injected each hour is defined as $P_{ren-i}$ and the power injected by conventional generation is $P_{gencon-i}$. The non-dispatchable load is $P_{load-nd-i}$, the flexible load which can be controlled is $P_{load-dis-i}$ and the total load $P_{load-tot-i}$. In order to make a fair assessment it is considered that in the period of time considered, the dispatchable load needs to supply a total amount of energy $E_{disp-tot}$. The different options are considered and compared:
\begin{itemize}[leftmargin=*]
    \item A - Constant dispatchable load and renewable generation operating with reserve $\alpha$ (to have the flexibility needed in grid-forming mode)
    \item B - Variable dispatchable load and renewable generation operating with reserve $\alpha$ (to have the flexibility needed in grid-forming mode)
    \item C - Grid forming load concept: Variable dispatchable load and renewable generation operating at the point needed by the operator allowing maximum generation.
\end{itemize}

The three options are analyzed formulating an optimization problem which minimizes the objective function
\begin{equation}
\text{Min } f = \sum P_{gencon-i}  
\end{equation}
which minimizes the total conventional generation adjusting the decision variables $P_{ren-i}$, $P_{load-dis-i}$, $P_{gencon-i}$ for all $i \in (1..h)$, subject to the constraints
\begin{eqnarray}
\sum P_{load-dis-i} &=& E_{disp-tot} \\
P_{ren-i} & \leq & \alpha P_{renav-i}      \\
P_{ren-i} + P_{gencon-i} &=&  P_{load-dis-i}  + P_{load-nd-i}  
\end{eqnarray}
for all $i \in (1..h)$. The formulation also ensures that all the variables and their variations are appropriately limited.  The optimization basically finds what is the best possible operation to minimizes conventional generation energy supplied during the period of analysis using an appropriate mix of renewable generation, dispatchable demand and conventional generation.

\figurename{~\ref{ener1}} shows the obtained results for the 3 options in a system with very high penetration of solar and wind analyzed for 150 hours. The upper Subfigure shows option A, where it can be noted the limitation on flexibility of demand is causing to reduce the renewable generation to 52.3 GWh requiring 6 GWh of conventional generation. The flexibility of the demand in option B (in the middle of \figurename{~\ref{ener1}}) is allowing to increase renewables injection to 54.4 GWh, while reducing significantly the conventional generation to 3.9 GWh. The grid-forming load concept proposed in option C (bottom of \figurename{~\ref{ener1}}) allows to improve even more the renewable generation, as it allows to operate at maximum power, thus bringing the renewables injection to 58.25 GWh and reducing the conventional generation to 0.11 GWh. It is important to remark that in all cases the total demand supplied (both flexible and non-dispatchable) is exactly the same.

\vspace{-3mm}
\section{Operation and Control Modes}\label{sec_controlmodes}

The GFM-L can operate in different operating modes depending on the overall state of the system. The following subsections address the main operation modes where the GFM-L can operate. In the normal operation mode, the GFM-L behaves as a GFM unit. Whenever there is a fault in the AC grid, the GFM-L will operate in fault operation mode. If the system is stopped and a black-start is needed, the GFM-L can be coordinated with a nearby renewable generation unit to facilitate network energizing.

\vspace{-3mm}
\subsection{Normal Operation Mode}

\begin{figure*}[htbp]
\centering
\includegraphics[width=.78\linewidth]{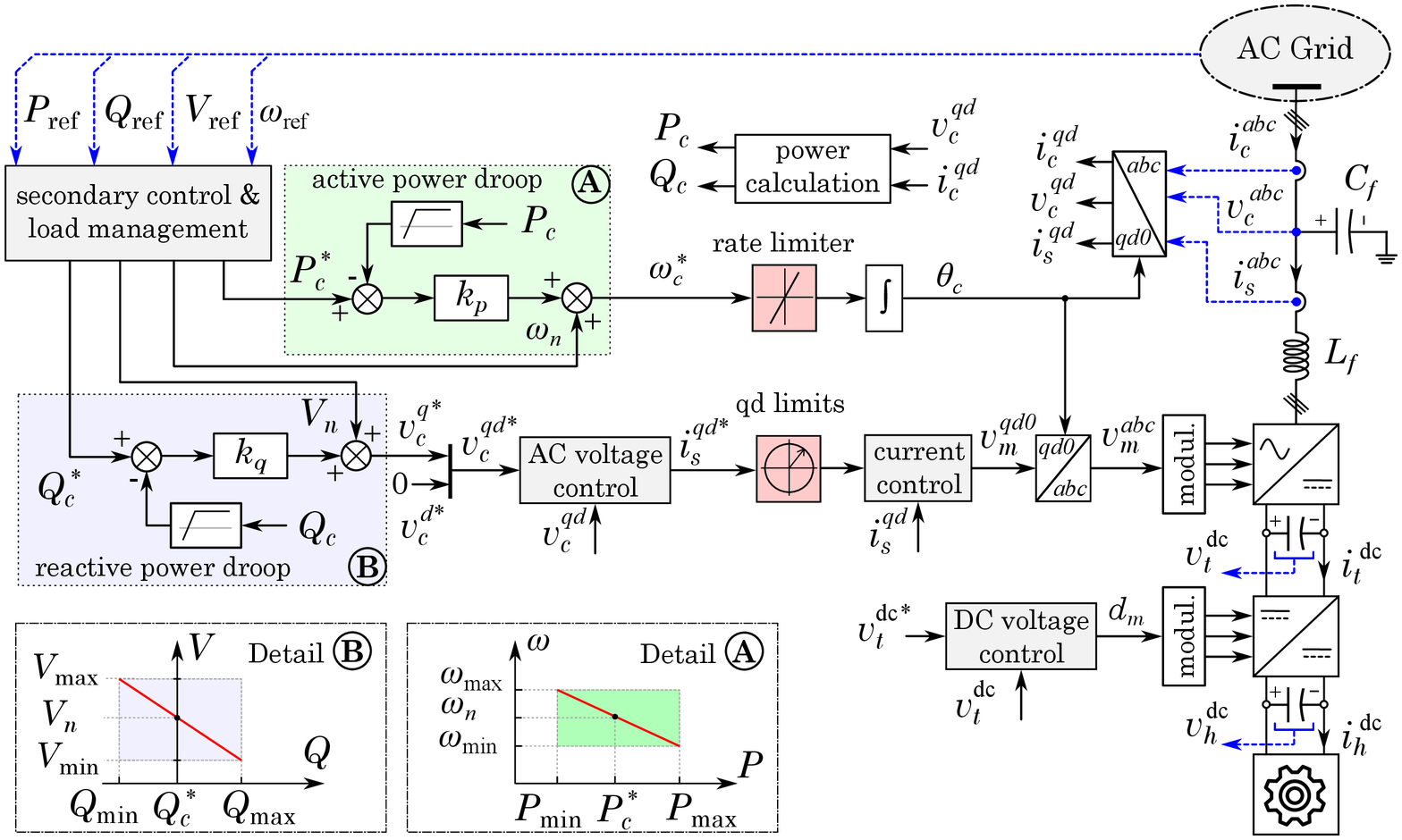}
\caption{Control system of a grid-forming load}
\label{GFM_control}
\end{figure*}

Without loss of generality, the control system of a GFM-L is shown in \figurename{~\ref{GFM_control}}. While a specific implementation is suggested in this paper, there can be other implementations using other grid-forming principles which could be adopted as well. The control system is based on the cascaded control method in which the slower outer control loops generate references for the faster inner control loops. By adopting such control strategy, a GFM-L behaves as a flexible voltage source which absorbs active power from the grid. The magnitude and angle of this voltage source are varied automatically (based on GFM law) depending on the variations in the voltage and frequency of the grid, which in turn changes the active/reactive power of the GFM-L. The major control blocks of such control system are as follow:
 
\begin{itemize}[leftmargin=*]
    \item \emph{Current control}: this control block regulates the GFM converter's output current, $i_s^{qd}$, at the references sent by the outer voltage control. The output of the current control loops, $v_m^{qd0}$, is transformed to the abc-frame via angle $\theta_c$ to build the modulation signal. Commonly, the current control has a time response of a few milliseconds.
    \item \emph{AC voltage control}: the outer AC voltage control regulates GFM converter's output voltage, $v_c^{qd}$, at the references produced by the reactive power droop. A standard qd-saturation block is placed on the output of the voltage control to protect the converter from overcurrent conditions. The response of the voltage control is at least ten times slower than that of the current control. 
    \item \emph{Active power droop}: the GFM-L is synchronized with the grid via this control block. A droop gain, $k_p$, is used to link active power deviation to the frequency deviation. Hence, if the grid's frequency is lower or higher than the nominal frequency ($\omega_{n}$), the power consumption of GFM-L is reduced or increased, respectively, based on the droop gain. The droop output, which is frequency reference, $\omega_{c}^*$, is passed through an integrator to build the angle of the voltage. A low-pass filter is placed on the feedback from the measured active power to emulate virtual inertia in the control system. The droop equation is given as
 \begin{equation}
    \omega_c^*=k_p~(P_c^*-H_p(s)~P_c)+\omega_n
    \label{P-f}
\end{equation}
where $H_p(s)$=$1/(\tau_p s+1)$ is a low-pass filter with time constant of $\tau_p$.
    \item \emph{Reactive power droop}: similar to the active power droop, a droop gain, $k_q$, is used to generate voltage reference in q-axis based on the reactive power deviation. Hence, if the grid's voltage goes higher or lower than the rated value, the GFM-L absorbs or injects more reactive power for compensation. The voltage reference in d-axis is set to zero. The peak of the base AC voltage is given by $V_n$. The reactive power droop is expressed by
\begin{equation}
    v_c^{q*}=k_q~(Q_c^*-H_q(s)~Q_c)+V_n
    \label{Q-v}
\end{equation}
where $H_q(s)$ is a low-pass filter with time constant of $\tau_q$ used to regulates the speed of voltage support. 
    \item \emph{DC voltage control}: the DC-link voltage of the GFM converter, $v_t^{\textrm{dc}}$, is tightly regulated at the nominal value via the DC voltage control loop of the DC-DC converter (or DC-AC converter in drive applications). The DC voltage at the terminal of the flexible load, $v_h^{\textrm{dc}}$, is allowed to deviate within a permissible range to provide flexibility in power consumption. It can be noticed that inversely to the conventional approach, in this case the load is the one responsible to control the DC voltage. So, the grid-side converter changes the power exchange according to the grid-forming control law and the load is the one in charge of regulating the DC voltage. 
    \item \emph{Secondary control and load management}: the grid operator can send a set of commands to the GFM-L, such as desired voltage magnitude ($V_{\textrm{ref}}$), frequency ($\omega_{\textrm{ref}}$), active and reactive power set points ($P_{\textrm{ref}}$ and $Q_{\textrm{ref}}$, respectively), to support grid's dynamics in different conditions. The secondary control receives those commands from the grid, and generates references for the active and reactive power droop controllers. The response of the secondary control is very slow as compared with that of outer voltage control. Moreover, a load management control is required to ensure that the operation condition of GFM converter is compatible with the mechanical/electrical constraints of the load. Such constraints depend on the load type (e.g industrial motor drives, in which temperature and mechanical loads of the motor may limit power and frequency of the GFM converter). 
\end{itemize}

The PI controller of the inner current control is tuned based on the filter impedance of the GFM converter. The proportional ($k_{pc}$) and integral ($k_{ic}$) gains are defined as
\begin{equation}
k_{pc}=\frac{L_f}{\tau_s} ~~,~~  k_{ic}=\frac{R_f}{\tau_s}
\end{equation}
where $L_f$ and $R_f$ are the filter inductance and resistance, and $\tau_s$ is the desired time constant (a few milliseconds). The PI controller of the outer voltage control loops can be defined as
\begin{equation}
k_{pv}=2  C_f  D_v \omega_v ~~,~~  k_{iv}=\omega_v^2  C_f
\end{equation}
where $D_v$ is the desired damping, commonly 0.707, and $\omega_v$ is the bandwidth of the voltage control loop, which is defined as 10$\times$2$\pi$/$\tau_s$ to be ten times slower than the inner current loops.

The tunings of the droop gains in the active and reactive power loops, $k_p$ and $k_q$, respectively, are based on the maximum permissible deviations in the frequency and voltage,
\begin{equation}
    k_p=0.02  \omega_n/S_n ~~,~~ k_q=0.1 V_n/S_n
\end{equation}
where $\omega_n$, $V_n$, and $S_n$ are the nominal frequency, voltage, and power of the GFM-L, respectively. Here, it is assumed that a maximum of $2\%$ of frequency and $10\%$ of voltage deviations are allowed during transitions. Nonetheless, the secondary controller can always bring the system back to the nominal frequency and voltage after disturbances.

The GFM-L is allowed to operate within active power consumption range [$P_{\textrm{min}}$,~$P_{\textrm{max}}$] as given by Detail~A in \figurename{~\ref{GFM_control}}. The actual power consumption depends on the frequency of the grid and the load mechanical/electrical constraints as defined in the secondary and load management control block. In addition, the GFM-L can regulate reactive power, independently from the active power consumption, within the range [$Q_{\textrm{min}}$,~$Q_{\textrm{max}}$] as given by Detail~B in \figurename{~\ref{GFM_control}}. However, the actual inductive/capacitive reactive power exchange with the grid depends on the voltage magnitude. Based on this, the  constraints during steady-state operation and transient responses are considered for a GFM-L $\omega_{\textrm{min}} \leq \omega_c \leq \omega_{\textrm{max}}$, $V_{\textrm{min}} \leq V_c \leq V_{\textrm{max}}$, $P_{\textrm{min}} \leq P_c \leq P_{\textrm{max}}$, $ Q_{\textrm{min}} \leq Q_c \leq Q_{\textrm{max}}$. The constraints are determined by the load's operational conditions and AC grid's requirements, which are integrated into the GFM-L control system through the secondary control and load management control block. $P_{\textrm{min}}$ can be regarded as the minimum power needed to keep the load operational. For example, it can be the minimum power required for a water pump system with frequency drive to displace steadily a minimum water flow. $P_{\textrm{max}}$ can be equal to the rated power consumption of the load. $Q_{\textrm{min}}$ and $Q_{\textrm{max}}$ are defined by the PQ capability curve of the GFM converter.


\subsection{Black-start Operation Mode}

The GFM-L can be utilized during black-start operation. Note that GFM-L alone cannot initiate the black-start operation and energize grid since it is a load in nature and lacks prime mover. However, in tandem operation with a nearby renewable generation system, it can maximize the benefits from renewable sources. 

One possible black-start sequence is that the renewable generation (in GFM mode) and GFM-L are started-up by ramping up their voltage references. During this period, active/reactive power loops are disconnected. Once voltage and frequency of the system reaches to the rated values, the active/reactive power loos are engaged. At this moment, both GFM renewable and GFM-L are participating in forming the grid. Next, the renewable generation can be switched to MPPT/grid-following operation to extract the maximum available renewable energy, while the GFM-L adjusts its active power consumption and reactive power exchange with the system to maintain voltage and frequency of the grid. Gradually, the rest of the network can be energized. 


\vspace{-3mm}
\subsection{Fault Operation Mode}

The GFM-L has to be protected during faults in the AC grid, as they may cause significant frequency deviation and overcurrent condition. To this aim, (i) a rate limiter is placed on the frequency reference to limit rate-of-change-of-frequency (ROCOF), commonly to 4 Hz/s during major transients, and (ii) a qd-saturation block is placed on the current reference to protect the GFM converter from overcurrent conditions. 

\begin{figure}[htbp]
\centering
\includegraphics[width=3in]{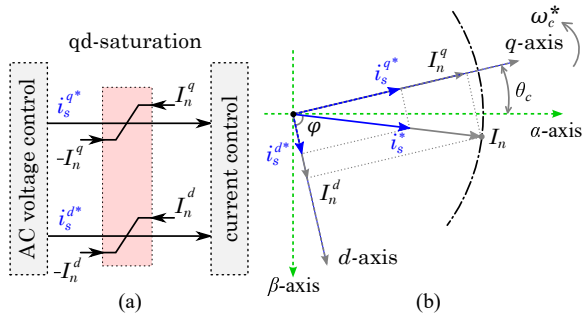}
\caption{Current saturation in qd=frame}
\label{saturation_limits}
\end{figure}

During fault condition in the AC grid, the magnitude of PCC voltage ($v_c^{abc}$ in \figurename{~\ref{GFM_control}}) dips to a very low value. The PI controller of the AC voltage loop tries to bring the PCC voltage back to its nominal value by increasing reactive current reference. Note that the GFM-L is not able to inject any active current to the grid since it is a load by nature. However, such reactive current injection may exceed the converter's current rating. Hence, the qd-current saturation block is used to limit the magnitude of the GFM's current reference, $i_s^{qd*}$, while keeping its angle constant in qd-axis. The upper and lower saturation limits are set dynamically. Assuming the rated current of the GFM converter is $I_n$, 
and referring to \figurename{~\ref{saturation_limits}},
\begin{equation}
    I_n^{q}=I_n |\sin{\phi}|
    \text{ , }
    I_n^{d}=I_n |\cos{\phi}|
    \label{In_qd}
\end{equation}
where $\phi$ is the angle of the current with respect to d-axis and simply calculated as $\phi= \arctan~ \frac{i_s^{q*}}{i_s^{d*}}$.
The positive values of (\ref{In_qd}) define the upper limits of current, and the negative values set the lower limits.

\vspace{-3mm}
\section{Case Studies}\label{sec_casestudies}

Four case studies are conducted in Matlab Simulink to verify the concept of GFM-L. In the first study (case study~1), a small network is used to discuss the fundamental advantages of the GFM-L. The second study deals with the black-start ability of GFM-L during grid energization. The third and fourth case studies aim at testing the performance of the GFM-L in a larger network with different penetration of renewable sources. 


\vspace{-3mm}
\subsection{Case Study~1: Essential System}

\begin{figure}[htbp]
\centering
\includegraphics[width=3.4in]{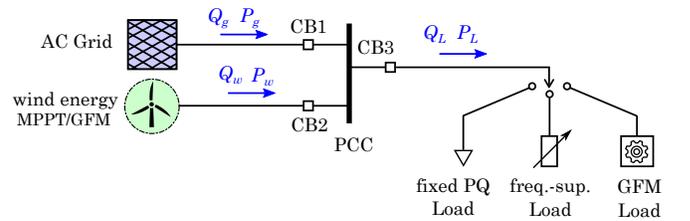}
\caption{Case study 1: the essential system topology}
\label{case_study_1}
\end{figure}

The network shown in \figurename{~\ref{case_study_1}} consists of a wind energy system, load, and AC grid. These three elements are essential to discuss the performance of GFM-L. The base power and voltage are 100~MV and 20~kV, respectively. The wind energy system can operate as grid forming to regulate voltage and frequency of the system, or as grid following to track the power reference generated by the maximum power point tracking (MPPT) control. The load can be any of these types: fixed-PQ load, frequency-support load, and GFM-L. The fixed-PQ load has no dependency on the frequency of the system, while frequency-support load has the capability of adjusting its active power consumption based on the system frequency, i.e., a power-frequency droop is used to generate power consumption from the network's frequency deviation. The GFM-L adopts the control system detailed in \figurename{~\ref{GFM_control}}.
The following cases are considered for performance analysis:
\begin{enumerate}[leftmargin=*]
    \item[(a)] wind energy is in MPPT mode and load is fixed-PQ type,
    \item[(b)] wind energy is in MPPT mode and load is frequency-support type,
    \item[(c)] wind energy is in GFM mode and load is fixed-PQ type,
    \item[(d)] wind energy is in MPPT mode and load is GFM type.
\end{enumerate}
The AC grid is based on the Th\'evenin equivalent model with a short circuit ratio of 2, and the equivalent inertia of 2~s. For all above-mentioned cases, the same simulation scenario is followed: initially, the system operates in grid-connected mode, where an under-frequency event occurs in the AC grid at t=2~s. Next, the grid disconnection (opening of CB1) occurs at t=3.6~s, leading to the island operation of the wind energy system and the load.

\begin{figure*}[htbp]
\centering
\includegraphics[width=0.9\linewidth]{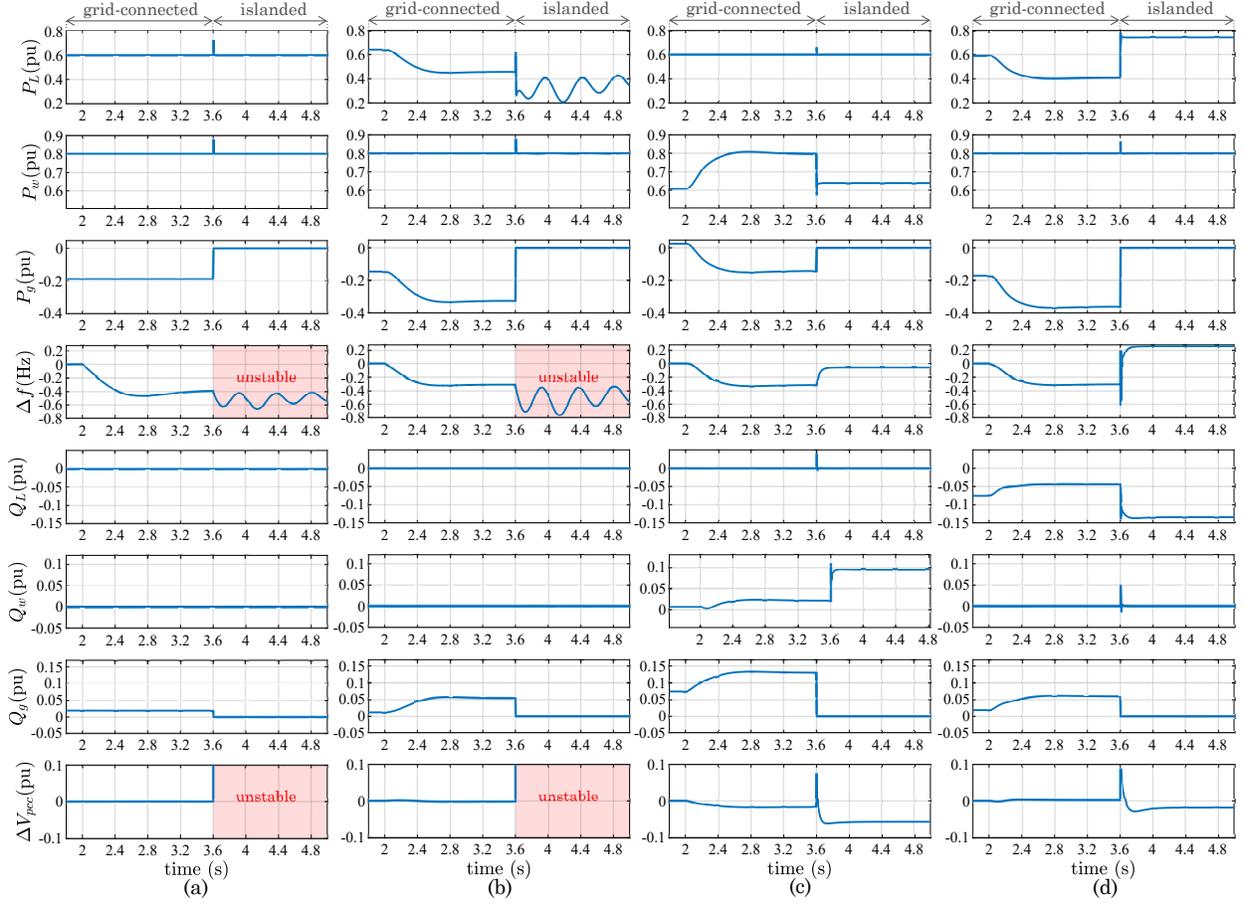}
\caption{Case study 1: (a) wind energy operates in MPPT mode with fixed-PQ load, (b) wind energy operates in MPPT mode with frequency-support load, (c) wind energy operates in GFM mode with fixed-PQ load, and (d) wind energy operates in MPPT mode with GFM-L}
\label{results_case_study_1}\vspace{-3mm}
\end{figure*}

The results obtained from the case~(a) are presented in \figurename{~\ref{case_study_1}(a)}. Since the load has a fixed-PQ type and the wind energy operates in MPPT mode (GFL mode), they do not provide support to the AC grid during under-frequency event. Hence, the load active power consumption, $P_L$, and the wind active power generation, $P_w$, remain unchanged before and after the frequency event at t=2~s. This situation is optimal for the wind energy since the maximum available power from the wind, assumed here to be 0.8~pu, is extracted and delivered to the load ($P_L$=0.6~pu) and the AC grid ($P_g$=0.2~pu); however, the frequency deviation, $\Delta f$, goes as low as -0.5~Hz during the transient. The main drawback of this configuration is that it cannot operate in islanded mode since neither the wind energy system nor the load has control provisions to form the voltage and frequency in the absence of the main AC grid. Therefore, the power imbalance between the wind and the load during islanded operation causes major frequency oscillations, and the point-of-common-connection (PCC) voltage deviation, $\Delta V_{pcc}$, goes beyond the limits. Eventually, case~(a) becomes unstable after grid disconnection. 

In the case~(b), similar system configuration is used, except that the load has the frequency-support provisions. Wind power is fixed at its MPPT value ($P_w$=0.8~pu), but $P_L$ depends on the frequency of the system. Hence, after the under-frequency event in the AC grid, $P_L$ is reduced from about 0.6~pu to 0.45~pu to allow the injection of more active power to the AC grid in response of the low frequency condition as shown in \figurename{~\ref{case_study_1}(b)}. Due to the frequency-support operation of the load, $\Delta f$ in the case~(b) is smaller than that of case~(a). This configuration provides support during grid's frequency disturbances, and the wind energy always operate at its maximum available power. However, similar to the case~(a), the case~(b) is also unable to regulate frequency and PCC voltage after AC grid disconnection, leading to an unstable system. 

In order to have a stable system during islanded operation, either wind energy system or the load has to operate in GFM mode. In the case~(c), the wind energy operates as GFM and the load has a fixed-PQ type. Although 0.8~pu power is available from the wind, the GFM operation requires lower power operation to ensure that the wind energy system is able to support the system during disturbances. As it is illustrated in \figurename{~\ref{case_study_1}(c)}, after the under-frequency event in the AC grid, $P_L$ remains unchanged, while $P_w$ is increased from 0.6~pu to 0.8~pu by the GFM operation due to the reduction in the system frequency. The frequency deviation is limited to about -0.4~Hz. After grid disconnection, the system continues stable operation as voltage and frequency are regulated by the GFM wind system. The main drawback of this configuration is that the maximum available power from the wind cannot be utilized since $P_w$ goes to only 0.5~pu to meet the load demand while 0.8~pu power is available from the wind energy. 

\begin{figure}[htbp]
\centering
\includegraphics[width=.48\textwidth]{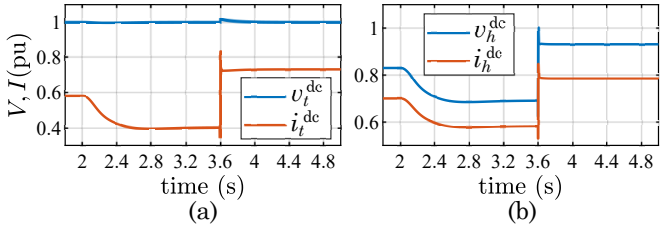}
\caption{Voltage and current of DC-side of GFM-L in the study case~(d): (a) DC link, (b) heater}
\label{results_case_study_1_2}\vspace{-3mm}
\end{figure}

Such drawback can be overcome by replacing fixed-PQ load with GFM-L and changing wind operation to MPPT mode. The results obtained from the case~(d) are presented in \figurename{~\ref{case_study_1}(d)}. The wind energy remains always at its maximum level, i.e., $P_w$=0.8~pu, regardless of the disturbances in the AC grid or grid disconnection. The GFM-L, however, reduces its active power consumption from 0.6~pu to 0.4~pu during under-frequency event. Here, a basic heater is assumed as the load. During the transients in the grid, the DC-DC converter located between the GFM converter and the load (please see \figurename{~\ref{GFM_control}}) tightly regulates the DC-link voltage, $v_t^{\textrm{dc}}$, at the nominal value; but it allows DC voltage at the heater terminal, $v_h^{\textrm{dc}}$, to deviate within a permissible range. The DC-link and heater voltages during the transient is shown in \figurename{~\ref{results_case_study_1_2}}. The heater can be regarded as a flexible load whose power consumption is linked to its DC voltage by $P_h$=$(v_h^{\textrm{dc}})^2$/$R_h$. The changes in the power consumption lead to variations in the heat generation of the heater. A summary of the advantages/disadvantages related to the case~(a) to (d) is provided in Table~(\ref{comp_table}).

\begin{table}
  \caption{Comparison table of case study~1}
  \label{comp_table}
  \includegraphics[width=0.9\linewidth]{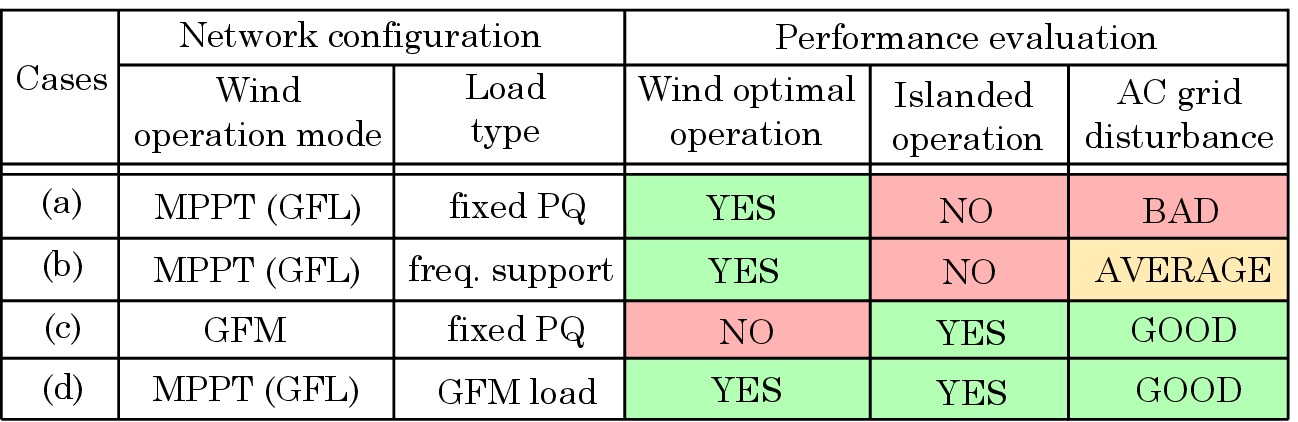}\vspace{-5mm}
\end{table}

\subsection{Case Study~2: Black-start Operation}
\begin{figure}[htbp]
\centering
\includegraphics[width=.48\textwidth]{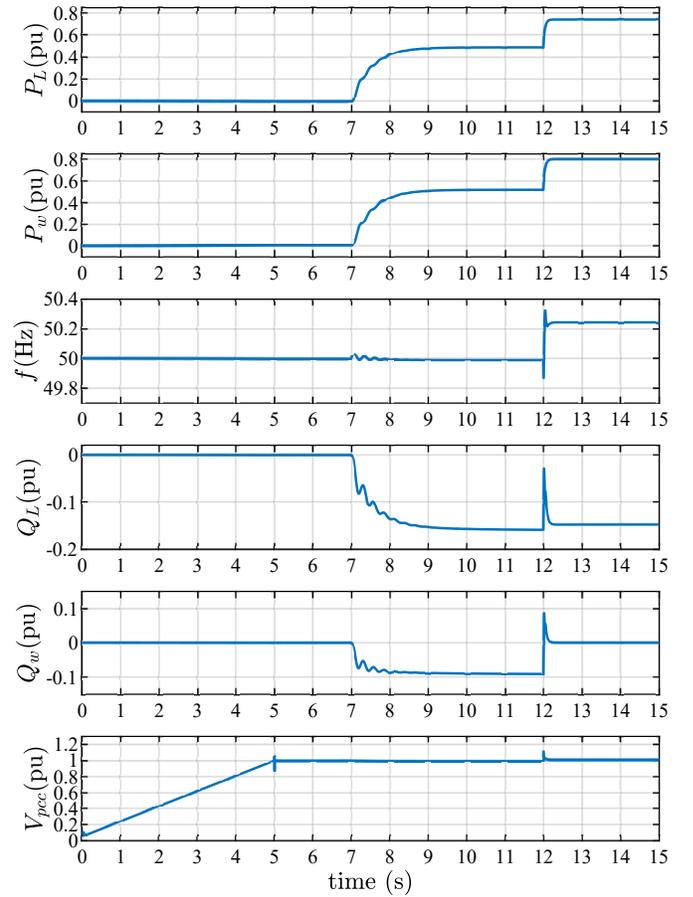}
\caption{Case study~2: black-start operation of the GFM-L in parallel with the wind energy}
\label{results_black_start}\vspace{-3mm}
\end{figure}

Assuming a blackout event occurs in the small system shown in \figurename{~\ref{case_study_1}}, the tandem operation of the wind energy and GFM-L can be utilized for black-start operation of the system. Following the black-start sequence explained in Section IV.B, the voltage references of the GFM-L and wind energy (operating in GFM mode) are ramped up until the system voltage reaches to its nominal value at t=5~s as shown in \figurename{~\ref{results_black_start}}. During the period of t=5 to 7~s, both wind energy and GFM-L participate in controlling the voltage and frequency of system, i.e., both operate in GFM mode. At t=7~s, the outer power loops are engaged and the power set-points are slowly increased, so that wind generation supplies the power demand of GFM-L, which is about 0.5~pu. At the final stage, the control system of the wind energy is switched to grid-following to track the maximum power point (0.75~pu), while GFM-L continues maintaining the voltage and frequency of system. The rest of the grid can be energized gradually. 

\vspace{-3mm}
\subsection{Case Study~3: Large Scale Integration of GFM-L With 100\% Penetration of Renewables}

\begin{figure*}[htbp]
\centering
\includegraphics[width=5.5in]{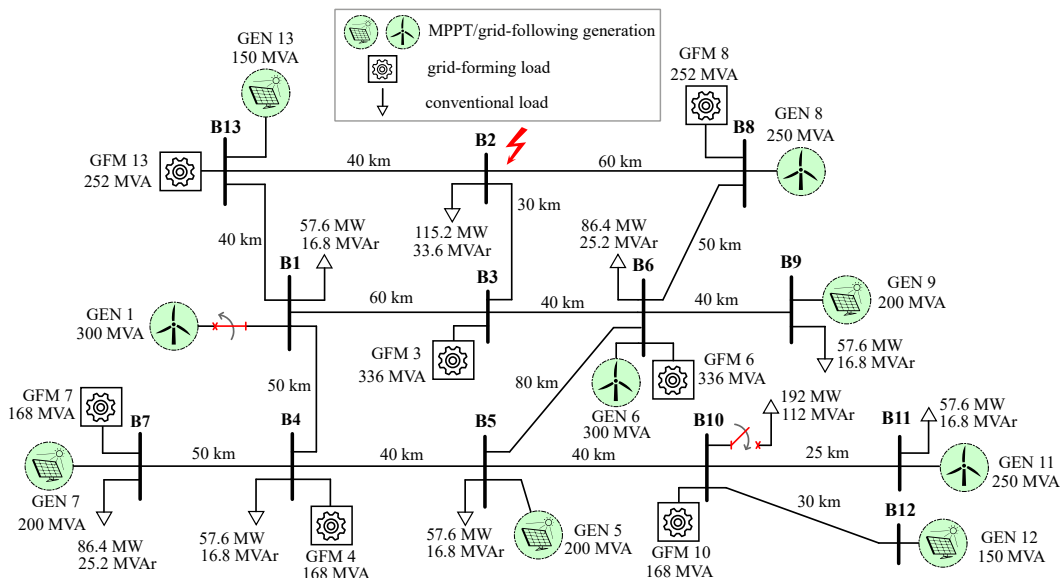}
\caption{Large system single-line diagram}
\label{large_system_diagram} \vspace{-5mm}
\end{figure*}

\begin{figure*}[htbp]
\centering
\includegraphics[width=5.5in]{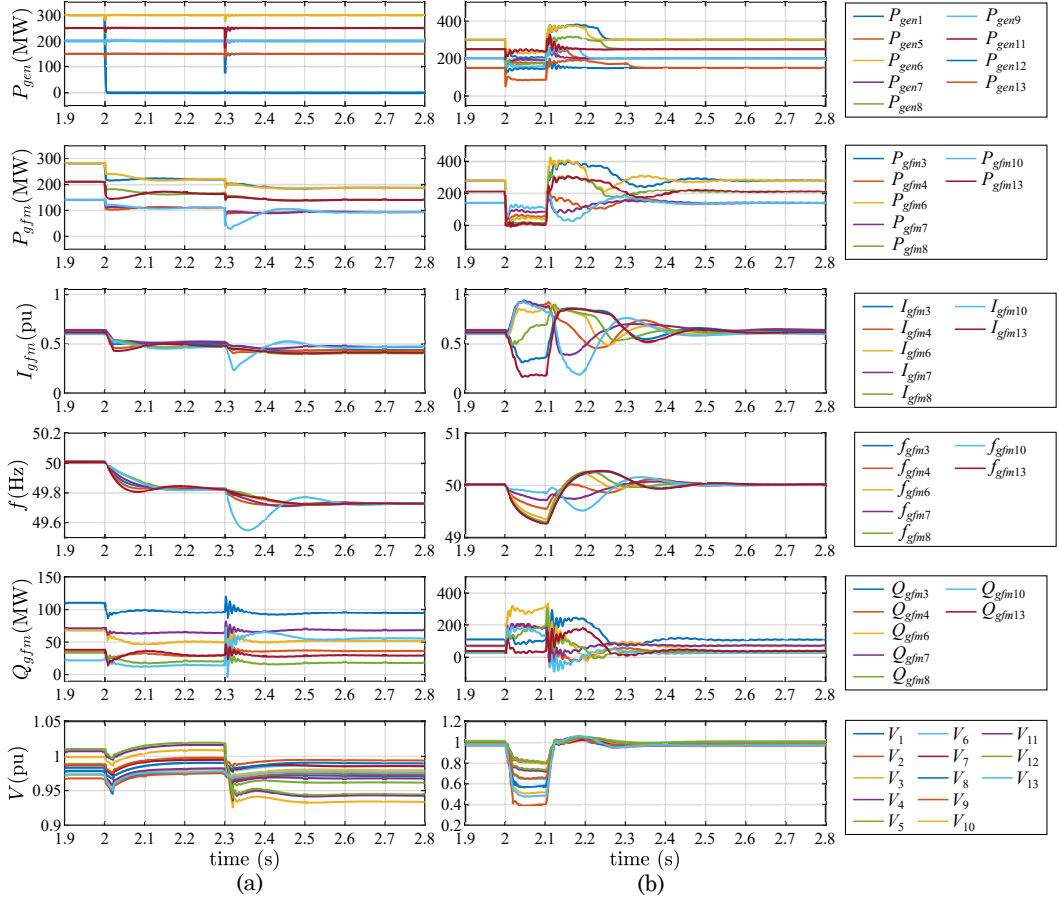}
\caption{Case study 3: (a) loss of power generation at t=2~s followed by load connection at t=2.3~s, (b) three-phase fault at t=2~s}
\label{results_case_study_2} \vspace{-5mm}
\end{figure*}

The dynamic behaviour of the GFM-L is tested in a larger power network whose single-line diagram is depicted in \figurename{~\ref{large_system_diagram}}. This exemplary power network is comprised of several renewable generators (wind and solar), conventional fixed-PQ loads, and GFM-L. Such power network represents a challenging futuristic power system with 100\% integration of renewable generations that all operate in MPPT/grid-following mode to ensure maximum extraction of available renewable energy. Note that the voltage and frequency regulations are entirely carried out by the GFM-L. The following cases are considered:
\begin{enumerate}[leftmargin=*]
    \item[(e)] the renewable generator that is connected to the bus B1 (indicated as GEN1 in \figurename{~\ref{large_system_diagram}}) is suddenly disconnected at t=2~s. This event is followed by the connection of a 192~MW, 112~MVAr fixed-PQ load to the bus B10 at t=2.3~s.
    \item[(f)] a three-phase fault occurs on the bus B2 at t=2~s, and it is cleared at t=2.1~s.
\end{enumerate}
The purpose of study case~(e) is to confirm the ability of GFM-L to overcome major generation and load changes in a large system. As it is illustrated in \figurename{~\ref{results_case_study_2}(a)}, after loss of Gen1 (300~MW) in the bus B1, $P_{gen1}$ drops to zero and the power consumption of all GFM-L, $P_{gfm}$, are reduced automatically to respond to this event. The generation loss causes frequency to drop to a lower value. Each GFM-L automatically reduces its power consumption to response to this under-frequency event in the system. The amount of power reduction in each GFM-L depends on its droop gain (see~(\ref{P-f})). Note that the rest of renewable generations continue operating at MPPT regardless of the disturbances in the network, so the maximum extraction of available renewable energy is ensured. Regarding the voltage profile throughout the system, it is maintained within $\pm10\%$ of nominal value during this transient.

Following to the previous event, A 192~MW/112~MVAr fixed-PQ load is connected to the bus B10 at t=2.3~s. Once again, this event causes another under-frequency event which results in further reduction in power consumption of GFM-L. Note that the frequency of the GFM-L connected to the bus B10, $f_{gfm10}$, has a larger deviation as it is located in close proximity of the load disturbance. Similar to the previous event, system remains stable and operates within permissible range.

The purpose of the study case~(f) is to confirm the ability of the network to ride through a three-phase fault by only relying on GFM-L. The fault occurs at t=2~s and cleared at t=2.1~s. The system performance is shown in \figurename{~\ref{results_case_study_2}(b)}. During the fault period, voltage of the bus B2, $V_2$, dips to 0.4~pu, and depending on the line impedances, the voltages of other buses also dip to a low magnitude. As a response to the fault condition, the active power consumption of the GFM-L are decreased significantly. The power consumption of those located in close proximity of the fault location, i.e., $P_{gfm3}$, $P_{gfm8}$, and $P_{gfm13}$, drop to nearly zero. On the contrary, reactive power of GFM-L are increased to support the voltage recovery. The RMS currents of the GFM-L during the fault are also illustrated in \figurename{~\ref{results_case_study_2}(b)}. The currents are in per units of the rated powers of the individual GFM-L. None of the currents exceed the rated current (1~pu) during fault. Once the fault is over, the system goes back to its stable operation. 

\subsection{Case Study~4: Large Scale Integration of GFM-L With 50\% Penetration of Renewables}

It is important to show how the proposed concept can be integrated with current power systems with a significant penetration of synchronous generators. In order to illustrate this, the network of case study~3 is used in this scenario, except that 50\% of renewables operating in MPPT/grid-following mode are replaced by synchronous generations. GEN5, GEN9, and GEN13 in \figurename{~\ref{large_system_diagram}} are replaced by traditional synchronous generators whose nominal powers add up to 50\% generation in the system (total generation is 2~GW, while 1~GW is coming from three synchronous generators). Moreover, the total power consumption in the system is divided equally between GFM loads and fixed-PQ loads. Under such consideration, the case study 3(e), which includes a generation loss and a large load connection (please see case study~3 for further details), is repeated here again, except that a longer simulation time is considered between the two events as synchronous generators have slower dynamics and need longer time to reach steady-state operation.

\begin{figure}[htbp]
\centering
\includegraphics[width=3.48in]{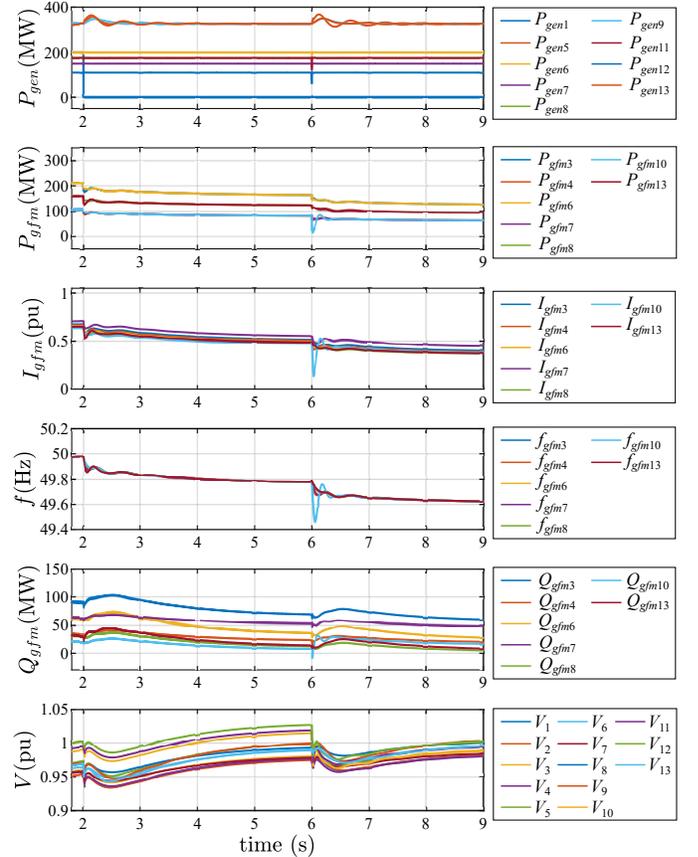}
\caption{Case study 4: loss of power generation at t=2~s followed by load connection at t=6~s}
\label{results_case_study_3} \vspace{-5mm}
\end{figure}

As illustrated in \figurename{~\ref{results_case_study_3}}, the generated power from the synchronous generators, $P_{gen5}$, $P_{gen9}$, and $P_{gen13}$, remain unchanged prior and after the generation loss (sudden disconnection of GEN1) at t=2~s. However, the power consumption of GFM loads are reduced as a response to such event. This is due to the fast control systems of GFM-L that are able to quickly respond to the disturbances. In fact, in this case, the GFM-L is designed to respond to power changes in only 40~ms. Comparing this to the inertia constant of the synchronous generators, which is $H=$5~s, it can be concluded that the GFM loads are reacting to power changes well prior to the slow synchronous generators. One advantage can be that in case a renewable source is out of service, and if there is enough flexibility in the GFM-L, the fast reduction in the power consumption of GFM-L can avoid synchronous generators to pick up the unnecessary portion of GFM loads and use fossil fuels to supply that. Such behaviour is repeated again once a 192~MW/112~MVAr fixed-PQ load is connected to the bus B10 at t=6~s. Again, GFM loads are the ones that quickly reduce their power consumption so that the new connected load is supplied with the existing available generation in the system and without increasing synchronous generator's power.

\section{Development Barriers}
\textcolor{blue}{As discussed throughout the paper, the development of the GFM-L concept offers some advantages for modern power systems. At the same time, there are some barriers which have to be considered to develop the concept further:
\begin{itemize}[leftmargin=*]
    \item The concept of GFM-L is applicable to the loads that can be dispatched and are equipped with a VSC. In some cases, the concept would require to replace the existing diode rectifier by a VSC in order to implement the concept.
    \item In systems or industries with multiple loads, the GFM-L can be applied locally and has an equivalent aggregated response from the grid perspective.
    \item The concept of GFM-L requires a load to operate based on the grid-forming law. Such law necessitates variations in the power consumption of a GFM-L according to the frequency condition in the grid (power-frequency droop control in GFM). However, there might be additional constraints on the load operation, such as load energy management optimization as discussed in~\cite{lian2022,xiao2021}, which also impose constraints on the power consumption of the load. It is important to make sure that there are no conflicts between grid-forming operation and other load constraints. 
    \item As loads can only absorb power, some critical power system operations (like black start) will require strong coordination between (variable renewable) generation and GFM-L. The paper has described and exemplified these cases but further work may be needed, also on establishing standards for interoperability as different equipment can be owned by different companies and supplied by different manufacturers.
    \item The previously mentioned coordination needs to be clearly defined in relevant grid codes, defining clear rules to ensure that responsibilities are properly assigned.
    \item Business models and economical aspects need to be further studied to establish fair regulation schemes to enable the GFM-L concept.
\end{itemize}}

\section{Conclusions}\label{sec_conclusions}
This paper proposed the concept of grid-forming loads as a solution to deal with the massive integration of renewables in modern power systems. Solar PV and wind power systems cannot store the resources and therefore it is convenient to operate them at the point of maximum power production. However, modern power systems need to be flexible and match generation and load trying to minimize the amount of energy storage needed. In this sense, the usage of demand management has emerged as a very important need in power systems worldwide. \textcolor{blue}{While significant work has been done on the area of demand response, this paper proposes for the demand-side to take a further step and provide grid-forming capabilities. Such concept is beneficial because 
\begin{itemize}[leftmargin=*]
    \item It can free renewable power plants from providing grid-forming services, and thus it allows them to operate at MPPT (in grid-forming mode renewable power plants need to operate with a reserve margin which reduces output power).
    \item It can guarantee safe operation of an islanded segment of power system while renewables can remain operating at MPPT (case study~1).
    \item It is able to help with the black-start sequence in which the tandem operation of a renewable source and grid-forming load is used to maximize renewable power during the black-start procedure (case study~2).
    \item It shows that the large integration of grid-forming loads can provide stability to a power system with 100\% penetration of renewable sources. The grid-forming capability is solely provided from grid-forming loads and all renewables are allowed to operate at MPPT (case study~3 and 4).
\end{itemize}}
To sum up, the main conclusion is that the grid-forming load is a promising concept and that further studies need to be done in the future for several specific types of loads, focusing on practical limitations that some loads can imply and considering these limitations in the overall design.
\bibliography{references}
\end{document}